\documentclass[12pt,fleqn]{article}
\usepackage{amssymb,cite}

\def\onecol{\onecolumn \mathindent 2em}

\def\noi{\noindent}

\makeatletter
\renewcommand{\section}{\@startsection{section}{1}{0pt}%
        {-3.5ex plus -1ex minus -.2ex}{2.3ex plus .2ex}%
        {\large\bf\protect\raggedright}}

\renewcommand{\subsection}{\@startsection{subsection}{2}{0pt}%
        {-3ex plus -1ex minus -.2ex}{1.4ex plus .2ex}%
        {\normalsize\bf\protect\raggedright}}

\renewcommand{\thesubsubsection}%
        {\arabic{section}.\arabic{subsection}.\arabic{subsubsection}.}

\renewcommand{\@oddhead}{\raisebox{0pt}[\headheight][0pt]{%
   \vbox{\hbox to\textwidth{\rightmark \hfil \rm \thepage \strut}\hrule}}}
\renewcommand{\@evenhead}{\raisebox{0pt}[\headheight][0pt]{%
   \vbox{\hbox to\textwidth{\thepage \hfil \leftmark \strut}\hrule}}}

\makeatother

\newcommand{\sect}[1]{Sec.\,#1}

\def\nqq{\hspace*{-2em}}
\def\nhq{\hspace*{-0.5em}}

\def\cm{\hspace*{1cm}}

\def\ten#1{\mbox{$\,\cdot\, 10^{#1}$}}

\def\Jl#1#2{{\it #1\/} {\bf #2},\ }

\def\ApJ#1 {\Jl{Astroph. J.}{#1}}
\def\CQG#1 {\Jl{Class. Quantum Grav.}{#1}}
\def\DAN#1 {\Jl{Dokl. AN SSSR}{#1}}
\def\GC#1 {\Jl{Grav. \& Cosmol.}{#1}}
\def\GRG#1 {\Jl{Gen. Rel. Grav.}{#1}}
\def\JETF#1 {\Jl{Zh. Eksp. Teor. Fiz.}{#1}}
\def\JETP#1 {\Jl{Sov. Phys. JETP}{#1}}
\def\JHEP#1 {\Jl{JHEP}{#1}}
\def\JMP#1 {\Jl{J. Math. Phys.}{#1}}
\def\NPB#1 {\Jl{Nucl. Phys.}{B\ #1}}
\def\NP#1 {\Jl{Nucl. Phys.}{#1}}
\def\PLA#1 {\Jl{Phys. Lett.}{#1A}}
\def\PLB#1 {\Jl{Phys. Lett.}{#1B}}
\def\PRD#1 {\Jl{Phys. Rev.}{D\ #1}}
\def\PRL#1 {\Jl{Phys. Rev. Lett.}{#1}}

\newcommand{\eqsection}{\makeatletter
    \@addtoreset{equation}{section}
    \renewcommand{\theequation}{\arabic{section}.\arabic{equation}}
    \makeatother}
\def\al{&\nhq}
\def\lal{&&\nqq {}}
\def\eq{Eq.\,}
\def\eqs{Eqs.\,}
\def\beq{\begin{equation}}
\def\eeq{\end{equation}}
\def\bear{\begin{eqnarray}}
\def\bearr{\begin{eqnarray} \lal}
\def\ear{\end{eqnarray}}
\def\earn{\nonumber \end{eqnarray}}

\def\eql{\al =\al}

\def\tst{\textstyle}

\def\fract#1#2{{\tst\frac{#1}{#2}}}

\def\half{{\fract{1}{2}}}

\def\e{{\,\rm e}}
\def\d{\partial}

\def\diag{\mathop{\rm diag}\nolimits}

\def\const{{\rm const}}

\def\then{\ \Rightarrow\ }

\def\R{{\mathbb R}}
\def\S{{\mathbb S}}

\def\mn{_{\mu\nu}}
\def\mN{_{\mu}^{\nu}}
\def\cA{{\cal A}}
\def\cM{{\cal M}}

\def\rhov{\rho_{\rm vac}}
\def\vac{{}_{\rm (vac)}}
\def\dens{\ {\rm g/cm^3}}

\begin{document}
\thispagestyle{empty}
\onecol

\begin {center}
{\Large\bf
  Regular homogeneous T-models with vacuum dark fluid}

\vskip0.2in

Kirill Bronnikov

{\sl VNIIMS, 3-1 M. Ulyanovoy St., Moscow 117313, Russia;}
{\sl Institute of Gravitation and Cosmology,
     PFUR, 6 Miklukho-Maklaya St., Moscow 117198, Russia}

\vskip0.1in

Irina Dymnikova

{\sl Department of Mathematics and Computer Science, University of
Warmia and Mazury,}\\
 {\sl \.Zolnierska 14,
    10-561 Olsztyn, Poland;}\\
{\sl A.F. Ioffe Physico-Technical Institute, Politekhnicheskaja 26,
St. Petersburg, 194021 Russia}

\end{center}

\begin{abstract}

We present the class of regular homogeneous T-models with vacuum dark fluid,
associated with a variable cosmological term. The vacuum fluid is defined by
the symmetry of its stress-energy tensor, i.e., its invariance under Lorentz
boosts in a distinguished spatial direction ($p_j=-\rho$), which makes this
fluid essentially anisotropic and allows its density to evolve. Typical
features of homogeneous regular T-models are: the existence of a Killing
horizon; beginning of the cosmological evolution from a null bang at the
horizon; the existence of a regular static pre-bang region visible to
cosmological observers; creation of matter from anisotropic vacuum,
accompanied by very rapid isotropization. We study in detail the spherically
symmetric regular T-models on the basis of a general exact solution for a
mixture of the vacuum fluid and dustlike matter and apply it to give
numerical estimates for a particular model which illustrates the ability of
cosmological T-models to satisfy the observational constraints.

\end{abstract}

\section{Introduction}

Astronomical data provide convincing evidence that our Universe is dominated
at above 70 per cent of its density by a dark energy responsible for its
accelerated expansion due to negative pressure, $p = w\rho$ with $w < -1/3$
\cite{riess,perlmutter,bahcall,spergel,shubnell}. Observations constrain the
equation of state of dark energy within $-1.45 < w < -0.74$ \cite{spergel},
the best fit  $w=-1$
\cite{scranton,corasaniti,hannestad,takahashi,tonry,ellis} corresponds to a
cosmological constant $\lambda$ related to the vacuum density $\rhov$ by
$\lambda = 8\pi G \rhov$ (for a detailed discussion see
\cite{spergel,ellis,copeland}).

The Einstein cosmological term $\Lambda g\mn = 8\pi G \rhov g\mn$ is plagued
with the {\it cosmological constant problem\/}. Quantum field theory
estimates $\Lambda$ by the Planck scale \cite{weinberg}, so that the
resulting zero-point density is incompatible with all observational data
\cite{krauss}, and a long-term problem was how to zero it out
\cite{weinberg}. A key problem now is why the cosmological constant should
be at the scale demanded by observations \cite{copeland}. Another dynamical
aspect is that the inflationary paradigm needs a large value of $\rhov$ at
the earliest stages to provide a reason for rapid initial expansion
\cite{gliner-dymnikova}, typically at the GUT scale \cite{guth} (for a
review see \cite{olive}), observations testify that it is small at the
present epoch, whereas the Einstein equations require $\rhov = \const$.

A lot of theories and models have been developed to describe dynamically
evolving dark energy which today behaves like a cosmological constant (for a
recent comprehensive review see \cite{copeland}).

On the other hand, the Einstein equations admit a class of solutions with
source terms associated with vacuum defined by the symmetry of its
stress-energy tensor, including vacuum with reduced symmetry which allows
the vacuum density to be evolving and clustering \cite{id2000,id2002}.

The Einstein cosmological term $\Lambda g\mn$ is associated with the
maximally symmetric vacuum stress-energy tensor (all eigenvalues equal)
$T\mn = \rhov g\mn$ \cite{gliner} representing de Sitter vacuum with
$p=-\rhov$ and $\rhov = \const$. The symmetry of a vacuum stress-energy
tensor can be reduced to the case when only one or two of the spacelike
eigenvalues of $T_{\mu\nu}$ coincide with its timelike eigenvalue,
$T_j^j=T_t^t$ \cite{id92,id2000,dg}, so that the vacuum equation of state
holds in the $j$-th direction(s), $p_j=-\rho$, and thus a vacuum with
reduced symmetry must be evidently anisotropic. Evolution of vacuum density
and pressures is governed by the conservation equation
$T^{\mu\nu}_{~~;\nu}=0$.

The class of GR solutions specified by $ p_j=-\rho$, which describe
time-dependent and spatially inhomogeneous vacuum energy, presents,  in a
model-independent way, {\it anisotropic vacuum fluid} with continuous
density and pressures. It can provide a unified description, based on a
space-time symmetry and introduced in \cite{dg}, of dark ingredients in the
Universe by a vacuum dark fluid representing both distributed vacuum dark
energy (by an evolving and inhomogeneous cosmological term \cite{id2000,bdd})
and compact self-gravitating objects with  interior de Sitter vacuum
\cite{id96,id2002,id2004,id2006}.

The variable cosmological term $\Lambda\mn = 8\pi G T\mn\vac$ has been
introduced in the spherically symmetric case \cite{id2000}~when  a vacuum
$T\mn\vac$ is invariant under radial Lorentz boosts \cite{id92}, while the
full symmetry is restored asymptotically at the centre and at infinity
\cite{id92,id96,id2002,ds}.

It provides a description of a smooth evolution of the vacuum from $\Lambda
g\mn$ to $\lambda g\mn$ with $\lambda < \Lambda$. In the spherically
symmetric case, $T\mn\vac$ generates globally regular space-times with a de
Sitter center \cite{id2002}. Depending on the parameters and on the choice
of a reference frame, they describe localized objects with  de Sitter vacuum
trapped inside (nonsingular black and white holes
\cite{id92,id2000,ddfg,dg2005}, self-gravitating vacuum structures without
event horizons \cite{id96} called G-lumps \cite{id2002,dg}), and regular
cosmological models with variable vacuum density and pressures  considered
and classified in our previous paper \cite{bdd}.

In the cosmological context, the spatial direction $x^j$ in which the vacuum
symmetry of $\Lambda\mn = 8\pi G T\mn\vac$ is preserved, is necessarily
distinguished by the expansion anisotropy. In cosmologies guided by
$\Lambda\mn$, an anisotropic stage is generic \cite{ddfg,bdd}. A further
early evolution is the most transparent in homogeneous T-models, and this is
the question addressed in the present paper.

T-models are generically related to the T-regions found originally
by Novikov as essentially non-static regions beyond the horizon in
 Schwarzschild space-time \cite{igor} and specified in the
spherically symmetric case, $ds^2 = \e^{\nu(q,t)}dt^2-
\e^{\mu(q,t)} dq^2-r^2(q,t)d\Omega^2$,  by the invariant quantity
$\Delta = g^{ik}r_{,i}r_{,k}
 = \e^{-\nu}(\d r/\d t)^2 - \e^{-\mu} (\d r/\d q)^2$ \cite{nf,bkt}.
When $\Delta > 0$, the normal $N_i=r_{,i}$ to the surface
$r=\const$ is timelike, so that the surface $r=\const$ is
spacelike, and orbits of constant $r$ and hence static observers
do not exist in principle. T-regions are further specified as
$T_-$ when $\dot{r} < 0$  and $T_+$ when $\dot{r} > 0$ (with $\Delta
> 0$, $\dot{r}$ cannot vanish, and this condition is
invariant)\cite{bkt}.\footnote
    {There are exceptions from this rule, represented by T-regions with
    $r(t)$ having a minimum. This happens, e.g., in regular black-hole
    solutions of brane gravity \cite{bra-bh} and in some recent
    solutions with nonminimally coupled Yang-Mills fields \cite{balak}.
    }

Formally, spherically symmetric T-models were introduced by Ruban
\cite{ruban} as modification of vacuum T-regions to the case of dustlike
matter considered in a synchronous comoving reference frame with the metric
representing an inhomogeneous generalization,
$ds^2=dt^2-e^{\alpha(t,q)}dq^2- R^2(t)d\Omega^2$, of the anisotropic
cosmological models of semi-closed type \cite{zeld} with hypercylindrical
spatial sections $V_3 = \S_2 \otimes \R_1$.  Actually Ruban deeply revised
and generalized (to the case of nonzero $\lambda$) Datt's solution
\cite{datt} qualified by the author as being ``of little physical
significance'' \cite{hans-juergen}. The particular case when
$\alpha(t,q)=\alpha(t)$ is known as the Kantowski-Sachs model \cite{kant}.
A remarkable feature of T-models discovered by Ruban is a gravitational
mass defect maximally possible in GR, equal to the total rest mass of the
matter which appears to be entirely gravitationally bound \cite{ruban}.

T-models were further extended to the case of a perfect fluid with non-zero
pressure \cite{ve,korkina,mw,shikin,collins}, and the  class  of
spherically symmetric T-models with isotropic perfect fluid was completed by
Ruban in 1983 \cite{ruban83}.

Amazingly, Ruban excluded from his consideration the case $T_0^0=T^1_1$
($p_r=-\rho$) as being ``of no interest for cosmology and relativistic
astrophysics'' \cite{thesis}. On the other hand, the requirement of
regularity leads in this case to globally regular space-times with an
obligatory de Sitter centre \cite{id2002} which represent distributed and
clustered vacuum dark energy and describe both black (white) nonsingular
holes and regular cosmological models dominated by anisotropic vacuum fluid,
including regular modifications of Kantowski-Sachs models \cite{bdd}. In
these models, the cosmological evolution starts from horizons with a highly
anisotropic null bang, but information on the pre-bang history in the de
Sitter region in the remote past is available to Kantowski-Sachs observers
\cite{bdd,beyond}.

The latter circumstance provides a solution to one more long-standing
problem, that of the initial cosmological singularity.

Homogeneous regular cosmological T-models with anisotropic vacuum fluid can
be seen as T-regions of space-times with 3-parameter isometry groups acting
transitively on spatial 2-surfaces, i.e., those with spherical, planar and
pseudospherical symmetries. Among them spherically symmetric T-models
related to Kantowski-Sachs (KS) type regular models are  of greatest
interest.

The aim of this paper is to introduce and to study spherically symmetric
homogeneous regular cosmological T-models with anisotropic vacuum fluid and
dustlike matter, bearing in mind that the material content of the modern
Universe is almost entirely represented by dark energy and pressureless
matter.

The paper is organized as follows. In \sect 2 we consider T-models of KS
type filled with dust and an anisotropic vacuum fluid described by
$\Lambda\mn$. For non-interacting dust and vacuum, we present a general
solution without specifying the particular density profile of $\Lambda\mn$.
In \sect 3 we argue that regular models on the basis of the above solutions
must have a static, purely vacuum, asymptotically de Sitter region in the
remote (pre-bang) past, and briefly describe the properties of such regular
models. In \sect 4 we give a detailed description of a particular model
illustrating the ability of regular T-models to satisfy the observational
constraints. In \sect 5 we summarize and discuss the results.

\section{General solution}

    We begin with the KS metric in the general form\footnote
{Our conventions are: the metric signature $(+{}-{}-{}-)$; the curvature
tensor $R^{\sigma}{}_{\mu\rho\nu} = \d_\nu\Gamma^{\sigma}_{\mu\rho}-\ldots,\
R\mn = R^{\sigma}{}_{\mu\sigma\nu}$, so that the Ricci scalar $R > 0$ for de
Sitter space-time and the matter-dominated cosmological epoch; the system of
units $c = \hbar = 1$.}
\beq                                                          \label{ds}
    ds^2 = \e^{2\gamma}dt^2 - \e^{2\alpha}dx^2
                    - \e^{2\beta}d\Omega^2
       \equiv  \e^{2\gamma}dt^2 - a^2(t) dx^2 -r^2(t) d\Omega^2
\eeq
    where $\alpha,\, \beta,\, \gamma$ are functions of the time coordinate
    $t$ and $d\Omega^2 = d\theta^2 + \sin^2 \theta d\varphi^2$. Then the
    nonzero Ricci tensor components are
\bear
    R^t_t \eql                                               \label{R00}
          \e^{-2\gamma}[\ddot{\alpha} + 2 \ddot{\beta}
                + \dot{\alpha}{}^2 + 2 \dot{\beta}{}^2
            - \dot{\gamma}(\dot{\alpha}+2\dot{\beta})];
\\
    R^x_x \eql                                           \label{R11}
          \e^{-2\gamma}[\ddot\alpha + \dot{\alpha}
                (\dot{\alpha}+2\dot{\beta}-\dot{\gamma})];
\\
    R^\theta_\theta = R^\varphi_\varphi \eql \e^{-2\beta}     \label{R22}
            + \e^{-2\gamma}[\ddot\beta + \dot{\beta}
                    (\dot{\alpha}+2\dot{\beta}-\dot{\gamma})]
\ear
    (the dot stands for $d/dt$).
    Two Einstein tensor components to be used are
\bear                                                        \label{G00}
    G^t_t \eql - \e^{-2\beta} - \e^{-2\gamma}
                  (\dot{\beta}{}^2 + 2\dot{\alpha}\dot{\beta}),
\\
    G^x_x \eql                                           \label{G11}
            - \e^{-2\beta} - \e^{-2\gamma}
         (2\ddot{\beta} + 3\dot{\beta}{}^2 -2\dot{\beta}\dot{\gamma}).
\ear

    The Einstein equations read
\beq                                                         \label{EE}
    G\mN \equiv R\mN - \half \delta\mN T^\rho_\rho = - 8\pi G T\mN.
\eeq

    For our purposes, it is helpful to choose the time coordinate $t$ in a
    way similar to choosing the curvature coordinates in static spherical
    symmetry, namely, to put $t=r \equiv \e^{\beta}$, so that the metric is
    rewritten as
\beq
    ds^2 = \e^{2\gamma(r)} dr^2 - \e^{2\alpha(r)}dx^2      \label{ds-r}
                        - r^2 d\Omega^2.
\eeq
    Then the ${1\choose 1}$ component of (\ref{EE}) and the
    difference ${0\choose 0}-{1\choose 1}$ read
\bear                                                          \label{E11}
      \frac{1}{r^2}\,
       \big[r(\e^{-2\gamma}+1)\big]' \eql 8\pi G T^x_x,
\\
      \frac{2}{r}\e^{-2\gamma} (\alpha' + \gamma')             \label{E0-1}
                   \eql 8\pi G (T^r_r - T^x_x),
\ear
    where the prime denotes $d/dr$. In full analogy with static spherical
    symmetry, \eqs (\ref{E11}) and (\ref{E0-1}) may be rewritten in an
    integral form:
\bear                                 \label{11int}
      \e^{-2\gamma} \eql -1 + \frac{8\pi G}{r}\int T^x_x\,r^2 dr,
\\
      \alpha  \eql - \gamma + 4\pi G                  \label{0-1int}
            \int r\e^{2\gamma} (T^r_r - T^x_x) dr.
\ear
    If $T^x_x$ is known as a function of $r$, then \eq (\ref{11int}) is a
    solution to (\ref{E11}) that yields the lapse function
    $\e^{2\gamma(r)}$, and the other metric function $\alpha$ is found from
    (\ref{0-1int}) if, in addition, the difference  $T^r_r - T^x_x$ is known
    as a function of $r$. It is really so in many important cases.

    Now, let us take the stress-energy tensor (SET) of matter as a sum
    of SETs of dust and vacuum. Dust, with the density $\rho_d(r)$, is
     at rest in a T-model with the metric (\ref{ds}), while for a spherically
     symmetric vacuum fluid  we have
    $T\mN{}_{\rm (vac)}= \diag (\rho_v,\ \rho_v,\ -p_{v\bot},\ -p_{v\bot})$.
    The full SET reads
\beq
      T\mN =                                                  \label{TmN}
           \diag (\rho_d + \rho_v,\ \rho_v,\ -p_{v\bot},\ -p_{v\bot}).
\eeq
    Evidently, if we know the vacuum density profile $\rho_v(r)$, \eq
    (\ref{11int}) gives us $\e^{2\gamma(r)}$:
\beq                                                          \label{gamma}
      \e^{-2\gamma} = -1 + \frac{2\cM(r)}{r},\cm
                       \cM(r) := 4\pi G \int \rho_v\,r^2 dr.
\eeq
    The function $\cM(r)$ is similar to the mass function conventionally
    introduced in the description of static, spherically symmetric
    configurations.

    Note that this result is obtained without using the conservation law
    for dust and vacuum and is even valid when they are interacting.

    It remains to find the dust density $\rho_d$ and the second metric
    function $\alpha(r)$. This can be done with the aid of the conservation
    law component $\nabla_\mu T^{\mu}_0 =0$ which in our case reads
\beq                                                            \label{cons}
     \rho'_d + \rho_d \biggl(\alpha' + \frac{2}{r}\biggr)
        + \rho'_v + \frac{2}{r} (\rho_v + p_{v\bot}) =0,
\eeq
    and \eq (\ref{E0-1}) where $T^r_r - T^x_x = \rho_d$. At stages of the
    cosmological evolution when there is no interaction between dust and
    vacuum, we can write the conservation law separately for each of them.
    Then, for dust we have
\beq                                      \label{cons-d}
     \rho'_d + \rho_d \biggl(\alpha' + \frac{2}{r}\biggr)=0
     \ \then \
        \rho_d = \mu_0 \e^{-\alpha}/r^2, \cm \mu_0 = \const,
\eeq
    while the corresponding relation for vacuum
\beq
    \rho'_v + \frac{2}{r} (\rho_v + p_{v\bot}) =0         \label{cons-v}
\eeq
    expresses $\rho_{v\bot}$ in terms of $\rho_v$. Substituting
    (\ref{cons-d}) into (\ref{E0-1}), we can rewrite it as follows:
\beq
      \e^{\alpha +\gamma}(\alpha' + \gamma') =
            \frac{4\pi G \mu_0}{r} \e^{3\gamma},            \label{0-1a}
\eeq
    which after integration gives
\beq                                                           \label{alpha}
      \e^{\alpha}
            = 4\pi G \mu_0\e^{-\gamma} \int \frac{dr}{r} \e^{3\gamma},
\eeq
    $\e^{\gamma}$ being given by (\ref{gamma}). This completes the solution.
    The physical time $\tau$ is obtained by integration,
\beq
    \tau = \int \e^{\gamma(r)} dr,                           \label{tau}
\eeq
    and all unknowns are expressed as functions of $\tau$ in a parametric
    form.

    Thus we have solved by quadratures the Einstein equations for KS
    cosmology with a non-interacting mixture of dust and
    $\Lambda\mn$-vacuum. The solution contains one essential integration
    constant $\mu_0$  and one arbitrary function, the vacuum
    density profile $\rho_v(r)$ (or equivalently the mass function
    $\cM (r)$). This is a new solution  for the general case
    $p_{v\bot} \ne -\rho_v$. In the particular case
     of the conventional cosmological constant $\lambda = 8\pi G \rho_v =
    -8\pi Gp_{v\bot} = \const$, our solution agrees with  Ruban's homogeneous
    solution \cite{ruban}.

    For interacting dust and vacuum, a further solution depends on the
    particular form of the interaction. Though, if we know (or have a reason
    to prescribe) the function $\rho_d (r)$, then \eq (\ref{0-1int})
    immediately yields $\alpha$ as a quadrature.

\section {Nonsingular models and the static core in the pre-bang past}

    Let us now discuss the possibility of obtaining globally regular models
    of the Universe on the basis of the above KS metrics. The following
    three points will be argued:
\begin{description}
\item[(i)]
    In regular expanding Universe models, the cosmological evolution
    should begin from a Killing horizon.
\item[(ii)]
    The region beyond the horizon should be purely vacuum, i.e.,
    dust is absent there.
\item[(iii)]
    Dust appears in the cosmological region due to interaction with the
    vacuum fluid.
\end{description}

    To prove item (i), we will use, in the metric (\ref{ds}), a new time
    variable specified by the coordinate condition $\alpha + \gamma =0$, so
    that (\ref{ds}) takes the form
    $ds^2 = a^{-2}(t) dt^2 - a^2(t) dx^2 - r^2(t) d\Omega^2$. Then, instead
    of (\ref{E0-1}), we obtain the equation
\beq
    2a^2 \frac{\ddot{r}}{r} = - 8\pi G \rho_d.        \label{r-ddot}
\eeq
    Assuming $\rho_d \geq 0$, we immediately conclude that $\ddot{r} \leq 0$.
    Consequently, if there is expansion ($\dot{r} > 0$) at some instant
    $t_1$, we can assert that $r(t)$ reached zero at some finite earlier
    instant $t_s < t_1$. Meanwhile, $r=0$ is a curvature singularity of the
    metric (\ref{ds}).

    The only way out is to assume that the cosmological evolution started
    {\it later\/} than the would-be singularity, which is only possible if
    there was a Killing horizon at some $t_h > t_s$, where $a(t_h) = 0$.
    Then there is a hope to find a static regular region beyond such a
    horizon. This reasonable assumption is additionally justified by the
    consideration below concerning (ii) and by the fact that all related
    regular vacuum models without dust \cite{bdd} do have horizons and
    regular R-regions with a de Sitter core in their remote past.

    As to item (ii), there is a serious physical argument. If we imagine
    that cosmological dust is continued to the static region where the
    cosmological time $t$ becomes a radial coordinate and vice versa, then
    the density $\rho_d=T^t_{t(d)}$ is converted there to radial pressure,
    while the pressure $-T^r_r=0$ is converted to zero density, and we
    obtain quite an unplausible kind of matter with zero density but nonzero
    pressure.

    Moreover, it can be strictly shown that $\rho_d =0$ at the horizon.
    Indeed, at $t=t_h$, according to (\ref{r-ddot}), $\rho_d =0$ due to
    $a=0$, unless $\ddot{r} \to \infty$ as $t\to t_h$. But the latter
    opportunity is unacceptable for a horizon as a regular surface where
    $r^2(t)$ should be an analytic function\footnote
{The coordinate defined by the condition $\alpha + \gamma =0$ varies near a
horizon (up to a nonzero constant factor) like manifestly well-behaved null
coordinates of Kruskal type used for analytic continuation of the metric
\cite{vac1,cold}. This implies the analyticity requirement for $r^2(t)$.}.
    Thus $\rho_d = 0$ at the horizon, and it is natural to conclude that
    dust did not exist to the past of it (item (ii)).

    Returning to the cosmological region, we notice that non-interacting
    dust is incompatible with a horizon: it would then have an infinite
    density according to (\ref{cons-d}), $\rho_d \sim [a(t) r^2(t)]^{-1}$.
    Since $\rho_d =0$ at the horizon, $\rho_d >0$ at later times can only
    emerge from the vacuum fluid due to interaction --- and this is our item
    (iii).

    Let us now consider the region beyond $t = t_h$ filled with vacuum fluid
    and suppose that there is a regular centre $r=0$. Such space-times were
    discussed in detail in Ref.\,\cite{bdd}; they can be described by
    metrics of the form (\ref{ds-r}) with $\e^{-2\gamma} = \e^{2\alpha}
    =-A(r)$ (so that $r$ is a spatial coordinate while $x$ is a temporal
    one), i.e.,
\beq
     ds^2 = A(r)dt_*^2 - \frac{dr^2}{A(r)} - r^2 d\Omega^2,\label{ds-stat}
\eeq
    with $A(r)=-e^{-2\gamma}$  given by \eq (\ref{gamma})\footnote
{Note that the time coordinate $t_*$ of the static region has nothing to do
with the cosmological time in the metric (\ref{ds}):  it is rather related
to the spatial coordinate $x$.}.  Taking at the centre $A(0) = 1$, we can
write
\beq                                                          \label{A(r)}
      A(r) = 1 - \frac{2\cM(r)}{r},\cm
                       \cM(r) = 4\pi G \int_0^r \rho_v\,r^2 dr,
\eeq
    so that $\cM$ is the conventional mass function.

    The function $A(r)$ may have different number and orders of zeros
    (depending on the particular density profile $\rho_v(r)$), corresponding
    to Killing horizons in space-time \cite{bdd}.

    In any space-time with the metric (\ref{ds-stat}), the geodesics
    equations have the following first integral:
\beq                                       \label{geod}
     \biggl(\frac{dr}{d\tau}\biggr)^2 + k A(r)
                + \frac{L^2}{r^2} A(r) = E^2 ,
\eeq
    where $E$ and $L$ are the constants of motion associated with the
    particle energy and angular momentum, while the constant $k$ takes the
    value $k=1$ for timelike geodesics and $k=0$ for null geodesics; the
    affine parameter $\tau$ has the meaning of proper time along a
    geodesic in the case $k=1$. At a horizon $r=r_h$, $A(r)$ vanishes, and
    in case $E\neq 0$ one has $dr/d\tau \neq 0$ for all geodesics, whence
    it follows that $|\tau| < \infty$, irrespective of the order of the
    horizon.  For KS comoving observers whose trajectories are time lines in
    the T-region, $A=1,L=0, E^2=0$.  Therefore even ``slow'' observers, whose
    world lines coincide with these time lines, receive information from
    their infinitely remote past brought by particles following the geodesics
    (\ref{geod}). So, KS observers, even in the case of a double horizon and
    a null bang in their infinitely remote past, can receive pre-bang
    information brought by particles and photons following other geodesics
    and crossing the horizon at their finite proper time.

    Let us here discuss the simplest generic structure with a single simple
    (first-order) horizon, so that the global space-time structure is like
    that of de Sitter space. The horizon radius $r=r_h$ corresponds to $r_h
    = 2\cM (r_h)$ and can be found if $\rho_v(r)$ is specified.

    The static R-region between $r=0$ and $r=r_h$ represents a regular
    ``core'' in the remote past of an expanding KS universe in which the
    cosmological evolution starts with a Null Bang (i.e., contains a horizon instead
    of a singularity).

    It is important that \eq (\ref{gamma}) with $\cM(r)$ given in
    (\ref{A(r)}) is now valid for all values of $r$, in the whole
    space-time, whereas for the other metric coefficient $e^{2\alpha}$ there
    are different expressions: $\e^{2\alpha} = -A(r)$ for $r \leq r_h$,
    (compare (\ref{ds-r}) and (\ref{ds-stat})); furthermore, \eq
    (\ref{alpha}) holds for the cosmological epoch without interaction,
    and there is a yet unknown expression for $e^{-2\alpha}$ in the
    interaction epoch.

    Since the horizon is crossed at a purely vacuum stage, a nonzero matter
    density $\rho_d$ appears due to interaction between dust and vacuum in a
    certain period after the Null Bang. This perfectly conforms to the
    well-known prediction of QFT in curved space-time that nonstationary
    space-times, especially anisotropic ones, create particles
    \cite{zeld,bird,gmm}, and a decaying vacuum density accompanied by
    growing $\rho_d$ is a phenomenological description of this process.

    In what follows we will give some estimates for a specific choice of the
    vacuum decay law.

\section{T-model with vacuum density
           $\rho_v = \rho_c \e^{-r^3/r_*^3} + \rho_\lambda$}

    Let us choose the profile $\rho_v(r)$ in the form
\beq                            \label{rho_v}
    \rho_v = \rho_c \e^{-r^3/r_*^3} + \rho_\lambda,
    \cm \rho_\lambda = \frac{\lambda}{8\pi G},
            \cm \rho_c,\ r_*, \lambda = \const >0,
\eeq
    The first term results from a simple semiclassical model for  vacuum
    polarization in spherically symmetric gravitational fields
    \cite{id92,id96,id2002}. The second one refers to a background
    cosmological term which is at the scale suggested by observations. An
    additional indirect justification of the ansatz (\ref{rho_v}) comes from
    the minisuperspace model of quantum cosmology. In a certain gauge, the
    cosmological constant is quantized as an eigenvalue of the appropriate
    potential in the Wheeler-DeWitt operator \cite{df}.  The form of the
    potential corresponds to a non-zero quantum value of the cosmological
    constant at each finite value of the time-dependent scale factor.

    We thus have three distinct length scales: $r_c = \sqrt{3/(8\pi
    G\rho_c)}$,  a de Sitter radius corresponding to the central density
    $\rho_c$, the scale $r_*\sim{(r_c^2 r_g)^{1/3}}$ \cite{werner,id92}
    characterizing the vacuum decay rate in the spherically symmetric
    gravitational field ($r_g$ is the Schwarzschild radius related to the
    total gravitational mass of the decaying vacuum),  and the scale
    $r_\lambda$ related to the background cosmological constant $\lambda$,
    assumed to be of the present Hubble order of magnitude, $\lambda \sim
    10^{-56}$ cm$^{-2}$, $r_\lambda \sim 10^{28}$ cm.

    Let us assume that the central density $\rho_c$ is of the GUT scale,
\beq
\label{rho_c}
     \rho_c \approx  (10^{15}\,{\rm GeV})^4 \approx 2.2\ten{75}\dens
     \ \then\
     r_c \approx  0.8 \ten {-25}\ {\rm cm},
\eeq
    and $r_* \approx  4.6\ten{-8}$ cm for the scale $r_{*}$ given
    in \cite{id92}.

\medskip\noi
{\bf Horizon radius.}
    Since $r_*\gg r_c$, evidently, the density $\rho_v$ is nearly constant
    and the metric is approximately de Sitter at radii $r \lesssim r_c$,
    and, with good accuracy, we have a cosmological horizon at $r \approx
    r_c$. The same is true for any profile $\rho_v(r)$ with a slow enough
    decay rate from $\rho_c$ corresponding to a regular centre. As a result,
    in all such cases, including (\ref{rho_v}),
\beq
         r_h \approx r_c,\cm  \rho_v (r_h) \approx  \rho_c.    \label{r_h}
\eeq
    More precisely, with (\ref{rho_v}) and (\ref{rho_c}), we have
    $r_h = r_c [1 + O(b^{-3})]$ where $b = r_*/r_c \approx 5.6 \ten {17}$.

\medskip\noi
{\bf Vacuum decay time.}
    Next, let us estimate the vacuum decay time, taking $\tau =0$ at
    crossing the horizon and seeking $\tau_f$, the instant when the
    first, exponential term in (\ref{rho_v}) becomes equal to the second
    term, $\rho_\lambda$. So,
\beq
      \tau_f = \int_{r_h}^{r_f} \e^{\gamma(r)} dr,             \label{tau_f}
\eeq
    where $r_f$ is found from the condition
\beq
    \rho_c \e^{-r_f^3/r_*^3} = \rho_\lambda \ \then\
        r_f \approx 6.3\,r_*.
\eeq
    The integrand $\e^{\gamma}$ is expressed from the relation due to
    (\ref{A(r)})
\beq
\label{2gam}
     \e^{-2\gamma} = -1 + \frac{b^2}{x} (1 - \e^{-x^3})
                           + \frac{\lambda}{3} r_*^2 x^2, \cm
        x: = \frac{r}{r_*}.
\eeq

    The integral (\ref{tau_f}) is taken over the interval $(1/b,\ 6.3)$. It
    is easily seen that, on this interval, the term containing $\lambda$ is
    more than 60 orders of magnitude smaller than the others and can be
    omitted.

    Near the horizon, where $x \ll 1$, the integral may be calculated
    analytically:
\[
     \tau_1 = \tau_f [r_1-r_h] \approx \int_{r_h}^{r_1}
            \frac{dx}{\sqrt{b^2x^2 -1}},
\]
    where $r_1$ is some small enough value of $r$. For $r_1 = 0.1\,r_*$
    we have (with a relative error smaller than $10^{-3}$) $\tau_1
    \approx 39.3\,r_c$.  In finding the remaining integral, we may leave in
    \eq (\ref{2gam}), with high precision, only the term with $b^2$ and get
    numerically
\[
     \tau_2 = r_c \int_{0.1}^{6.3}
         \frac{\sqrt{x}\,dx}{\sqrt{1-\e^{-x^3}}} \approx 12.3\, r_c,
\]
    which finally gives
\beq
     \tau_f = \tau_1 + \tau_2 \approx 51,6\,r_c            \label{tau-f}
                    \approx 0.4\ten{-23}\,{\rm cm},
\eeq
    or, in seconds, $\tau_f \approx 1.3\ten{-34}$ s. This value is rather
    close to the GUT parameters, while by this time the scale factor $r$
    has already inflated from $r = r_h \approx 0.8 \ten{-25}$ cm to
    $r_f\approx 6.3 b\,r_c \approx 2.8\ten{-7}$ cm. The other scale
    factor, $a(r)$, has inflated from zero at the horizon to some finite
    value.

    The inflation of $r$ corresponds to approximately 43 e-foldings which
    is not regarded sufficient in conventional inflationary cosmology.
    However, in our class of models inflation as such is not necessary
    because, due to the existence and observability of a static core in the remote past,
    all parts of our model Universe are causally connected.

\medskip\noi
{\bf Solution in the regime of constant $\lambda$.}
    In the period $\tau > \tau_f$ we can neglect the variable part of the
    vacuum SET, remaining with the usual cosmological term,
    $\rho_v = - p_{v\bot} = \lambda/(8\pi G) > 0$. Then, according to
    (\ref{gamma}), the lapse function is given by
\beq
\label{gamma2}
      \e^{-2\gamma} = - 1 + \frac{2M} {r} + \frac{1}{3}\lambda r^2,
\eeq
    where $M$ is an integration constant expressing the total contribution
    of the decaying vacuum component to the mass function (this corresponds
    to dropping $\e^{-x^3}$ in (\ref{2gam}), and we have then
    $2M = b^2 r_* \approx 1.4 \ten {28}$ cm).

    Integration in \eq (\ref{alpha}) then leads to a very cumbersome
    expression in terms of elliptic functions, and we will not present it
    here. We notice that, in the absence of dust, the integral in
    (\ref{alpha}) reduces to an arbitrary constant, and after simple
    rescaling of $t$ (or directly from \eq (\ref{E0-1})) we obtain the
    conventional Schwarzschild-de Sitter metric in its T region, i.e.,
    $\alpha = -\gamma$ with $\gamma$ given by (\ref{gamma2}).

\medskip\noi
{\bf Matter-dominated epoch.}
    In our model, there is a large period when
\beq
       r_f \lesssim r \ll M \sim 10^{28}\ {\rm cm},          \label{r-mat}
\eeq
    in which $\rho_d \gg \rho_v$ and we can write $\e^{2\gamma} \approx
    2M/r$. This corresponds to the epoch when all matter has been already
    created but the influence of the cosmological constant $\lambda$ is
    still negligible. In this period, the integral in (\ref{alpha}) is
    easily calculated, and both $r$ and $\alpha$ may be expressed in terms
    of the cosmic time $\tau$:
\bear
     a(r) = \e^\alpha = \frac{8\pi G \mu_0}{3M}              \label{a-mat}
                \biggl(r + \frac{c_1}{\sqrt{r}} \biggr),
\cm
     r =  \left[ 3\sqrt{M/2}\, (\tau-\tau_0)\right]^{2/3},
\ear
    where $c_1$ and $\tau_0$ are integration constants. This very simple
    expression gives a reasonable approximation to the corresponding
    exact dust solution \cite{ruban83}.

    The values of $c_1$ and $\tau_0$ are determined by the details of the
    interaction period $r < r_f$. If we admit for certainty that the
matter-dominated epoch begins at $r = r_f$  and, as before, put $\tau = 0$
    at the horizon, we must require $r(\tau_f) = r_f$, which leads to
    $\tau_f - \tau_0 \approx 0.8\ten{-24}$ cm, and using (\ref{tau_f}) we
    obtain $\tau_0 \approx 3.2\ten{-24}$ cm.

    The value of $c_1$ is severely restricted by the conditions $a >0$ and
    $a' > 0$ at $r > r_f$:
\beq
    -r_f^{3/2} < c_1 < 2\,r_f^{3/2}.                    \label{c1}
\eeq

\medskip\noi
{\bf Isotropization.}
    The condition that a KS universe expands isotropically is that the
    directional Hubble parameters, defined as \cite{craw}
\beq
    H_a := \e^{-\gamma} \dot \alpha \equiv d\alpha/d\tau,
       \cm  {\rm and} \cm
    H_r := \e^{-\gamma} \dot \beta \equiv d\beta/d\tau = (1/r)(dr/d\tau),
\eeq
    are equal. Entire isotropy would mean that $R^x_x = R^\theta_\theta$,
    which, according to (\ref{R11}) and (\ref{R22}), would lead (in the
    proper time gauge $t = \tau$) to
\[
    \ddot \alpha + 3 \dot\alpha{}^2 =
    \ddot \alpha + 3 \dot\alpha{}^2 + 1/r^2.
\]
    This shows that any KS cosmology is necessarily anisotropic for
    topological reasons and can only isotropize asymptotically at large $r$.

    The degree of anisotropy may be characterized by the ratio
    $\cA = \sigma/H$ where
\beq
        \sigma = (H_a - H_r)/\sqrt{3}, \cm H = (H_a + 2H_r)/3
\eeq
    are the shear scalar and the mean Hubble parameter, respectively.
    Since $H_a$ and $H_r$ can in principle be equal, the observable
    parameter $\cA$ can be zero despite the topological anisotropy.

    Observations of the cosmic microwave background show that our Universe
    is highly isotropic \cite{cmb}: at the recombination epoch, at which the
    electromagnetic radiation has decoupled from matter (at redshifts
    $z \approx 1000$), the ratio $\sigma/H$ could not exceed $10^{-6}$,
    whence it follows that at present it is at most of order $10^{-9}$
    \cite{craw}. This, in principle, strongly constrains the parameters of
    any cosmological model based on the KS metric. We shall see, however,
    that our model is automatically sufficiently isotropic.

    In terms of the quantities $r$ and $a(r)$, the anisotropy parameter
    reads
\beq
    \cA = \sqrt{3}\, \frac{r a'(r) -1}{r a'(r) + 2}.
\eeq
    For the matter-dominated epoch we obtain from (\ref{a-mat})
\beq
    \cA = - \frac{\sqrt{3} c_1}{2 r^{3/2} + c_1}.
\eeq
    From the restriction (\ref{c1}) it is evident that our model Universe
    expands almost isotropically very soon after the end of the interaction
    epoch, and at $r \gg r_f$ we have simply
\beq
    \cA \lesssim (r_f/r)^{3/2}.
\eeq
    For instance, in the recombination epoch ($r \sim 10^{25}$ cm), still
    belonging to the matter-dominated period, we obtain $\cA \lesssim
    10^{-48}$, a degree of isotropy far exceeding the observational
    constraint.

\medskip\noi
{\bf Matter density after creation.}
    It is reasonable to suppose that, by the time $\tau = \tau_f$, when the
    rapidly decaying vacuum component almost completely vanishes, the whole
    amount of matter particles had already been created, and the further
    evolution proceeded with each of the conservation laws (\ref{cons-v})
    and (\ref{cons-d}) for dust and vacuum being valid separately, hence we
    have the expression (\ref{alpha}) for $a(r) = \e^\alpha(r)$.

    Let us obtain an order-of-magnitude estimate of matter density at
    $\tau = \tau_f$, using the conservation law (\ref{cons-d}), assuming
    that at present, when we can put $r =a =r_0 \approx 10^{28}$ cm
    the dust density corresponds to the total (visible plus dark) matter
    density of the Universe, i.e.,
\beq
       \rho_d \approx m_0/r_0^3 \approx 2\ten{-30}\dens, \cm
             m_0 =\const \approx 2\ten{54}\ {\rm g}.
\eeq
    Neglecting the anisotropy (which, as we saw, is really negligible soon
    after $\tau = \tau_f$), i.e., assuming $a(r) \approx r$ at all
    $\tau > \tau_f$, we obtain
\beq
    \rho_d (\tau_f) \approx 0.8 \ten{74}\dens
                        \approx \frac{1}{2600}\ \rho_c.
\eeq
    This seems to be a reasonable figure for the state of the Universe right
    after the vacuum decay and matter creation.

\section{Conclusion}

A general solution for homogeneous T-models has been found for a mixture of
vacuum dark fluid and dustlike matter. It presents the class of T-models
specified by the density profile of the vacuum fluid $\rho_v$. The
solution contains one arbitrary integration constant related to the dust
density.

We have proved the existence of regular homogeneous T-models with the
following structure:

\begin{description}
\item[(i)]
    A regular static R-region in the pre-bang past, which is
    asymptotically de Sitter as $r\to 0$.

\item[(ii)]
    Killing horizon(s) separating static R-region(s) from KS T-region(s).

\item[(iii)]
    A null bang from a Killing horizon, followed by matter creation
    from an anisotropic vacuum, accompanied by rapid isotropization.

\item[(iv)]
    Further evolution in the regime without interaction between
    matter and vacuum.
\end{description}

Let us emphasize that the rapid isotropization was not imposed by hand. For
our regular T-models rapid isotropization is generic, since the cosmological
evolution starts from the Killing horizon, in this case $a>0, \dot{a}>0$
which immediately leads to tight constraints on the anisotropy factor.

The most interesting feature is that information about pre-bang history
is available, and hence observable for the comoving observers in T-regions
and can influence dynamics.

Let us also emphasize that regular homogeneous T-models with anisotropic
vacuum dark fluid do not not need an inflationary stage since there are no
causally disconnected spatial regions. Still, in this class of models, there
is a certain amount of inflation, which happens simultaneously with
isotropization, right after the Null Bang: the scale factor $a(\tau)$ grows
from zero to a finite value while the other scale factor $r(\tau)$ rapidly
grows from one finite value to another being driven by the (currently large
but decaying) variable $\Lambda$-term.

\vskip0.2in

{\bf Acknowledgement}

\vskip0.1in

This work was supported by the Polish Ministry of Science and Information
Society Technologies through the grant 1P03D.023.27.  K.B. was partly
supported by RFBR Project No. 05-02-17478. I.D. is grateful to Malcolm
MacCallum for the instructive brief e-mail discussion of T-models.

\end{document}